\documentstyle[epsf,psfig]{mn}

\newif\ifAMStwofonts

\ifoldfss
  \ifCUPmtlplainloaded \else
    \NewTextAlphabet{textbfit} {cmbxti10} {}
    \NewTextAlphabet{textbfss} {cmssbx10} {}
    \NewMathAlphabet{mathbfit} {cmbxti10} {} 
    \NewMathAlphabet{mathbfss} {cmssbx10} {} 
  \fi
  \ifAMStwofonts
    \ifCUPmtlplainloaded \else
      \NewSymbolFont{upmath} {eurm10}
      \NewSymbolFont{AMSa} {msam10}
      \NewMathSymbol{\upi}     {0}{upmath}{19}
      \NewMathSymbol{\umu}     {0}{upmath}{16}
      \NewMathSymbol{\upartial}{0}{upmath}{40}
      \NewMathSymbol{\leqslant}{3}{AMSa}{36}
      \NewMathSymbol{\geqslant}{3}{AMSa}{3E}

    \fi
  \fi
\fi 

\ifnfssone
  \newmathalphabet{\mathit}
  \addtoversion{normal}{\mathit}{cmr}{m}{it}
  \addtoversion{bold}{\mathit}{cmr}{bx}{it}
  \newmathalphabet{\mathbfit} 
  \addtoversion{normal}{\mathbfit}{cmr}{bx}{it}
  \addtoversion{bold}{\mathbfit}{cmr}{bx}{it}
  \newmathalphabet{\mathbfss} 
  \addtoversion{normal}{\mathbfss}{cmss}{bx}{n}
  \addtoversion{bold}{\mathbfss}{cmss}{bx}{n}
  \ifAMStwofonts
    \ifCUPmtlplainloaded \else
      \UseAMStwoboldmath
      \makeatletter
      \new@mathgroup\upmath@group
      \define@mathgroup\mv@normal\upmath@group{eur}{m}{n}
      \define@mathgroup\mv@bold\upmath@group{eur}{b}{n}
      \edef\UPM{\hexnumber\upmath@group}
      \new@mathgroup\amsa@group
      \define@mathgroup\mv@normal\amsa@group{msa}{m}{n}
      \define@mathgroup\mv@bold\amsa@group{msa}{m}{n}
      \edef\AMSa{\hexnumber\amsa@group}
      \makeatother
      \mathchardef\upi="0\UPM19
      \mathchardef\umu="0\UPM16
      \mathchardef\upartial="0\UPM40
      \mathchardef\leqslant="3\AMSa36
      \mathchardef\geqslant="3\AMSa3E
    \fi
  \fi
\fi 

\ifnfsstwo
  \DeclareMathAlphabet{\mathbfit}{OT1}{cmr}{bx}{it}
  \SetMathAlphabet\mathbfit{bold}{OT1}{cmr}{bx}{it}
  \DeclareMathAlphabet{\mathbfss}{OT1}{cmss}{bx}{n}
  \SetMathAlphabet\mathbfss{bold}{OT1}{cmss}{bx}{n}
  \ifAMStwofonts
    \ifCUPmtlplainloaded \else
      \DeclareSymbolFont{UPM}{U}{eur}{m}{n}
      \SetSymbolFont{UPM}{bold}{U}{eur}{b}{n}
      \DeclareSymbolFont{AMSa}{U}{msa}{m}{n}
      \DeclareMathSymbol{\upi}{0}{UPM}{"19}
      \DeclareMathSymbol{\umu}{0}{UPM}{"16}
      \DeclareMathSymbol{\upartial}{0}{UPM}{"40}
      \DeclareMathSymbol{\leqslant}{3}{AMSa}{"36}
      \DeclareMathSymbol{\geqslant}{3}{AMSa}{"3E}
    \fi
  \fi
\fi 

\ifCUPmtlplainloaded \else
  \ifAMStwofonts \else 
    \def\upi{\pi}
    \def\umu{\mu}
    \def\upartial{\partial}
  \fi
\fi

\title{AX J0049.4-7323 - a close look at a neutron star interacting
with a circumstellar disk}
\author[M.J. Coe et al.]
       {M. J.~Coe and W.R.T.Edge  \\
       School of Physics and Astronomy, Southampton University, SO17 
1BJ, UK\\}
\date{Accepted .
      Received ;
      in original form
      }

\pagerange{\pageref{firstpage}--\pageref{lastpage}}
\pubyear{2003}

\begin{document}

\bibliographystyle{plain}

\maketitle


\begin{abstract}

Detailed evidence on the system AX J0049.4-7323 is presented here to
show how the passage of the neutron star in the binary system disrupts
the circumstellar disk of the mass donor Be star. A similar effect is
noted in three other Be/X-ray binary systems. Together the
observational data should provide valuable tools for modelling these
complex interactions.

\end{abstract}

\section{Introduction and background}

The Be/X-ray systems represent the largest sub-class of massive X-ray
binaries.  A survey of the literature reveals that of 96 proposed
massive X-ray binary pulsar systems, 57\% of the identified systems
fall within this class of binary.  The orbit of the Be or supergiant
star and the compact object, presumably a neutron star, is generally
wide and eccentric.  X-ray outbursts are normally associated with the
passage of the neutron star close to the circumstellar disk (Okazaki
\& Negueruela, 2001). A recent review of these systems may be found in
Coe (2000).

The optical light from a Be/X-ray binary is dominated by the mass donor
star in the blue end of the spectrum, but at the red end there is
normally a significant contribution from the circumstellar disk. Long
term optical observations such as those collected by the MACHO
experiment (Alcock et al, 1995) provide valuable insights into the
behaviour of the circumstellar disk, and hence into some of the
details of the binary interactions within the system.

\section{AX J0049.4-7323}

\subsection{Optical and X-ray outbursts}

The optical counterpart to AX J0049.4-7323 (Yokogawa et al, 2000) 
was identified by Edge \&
Coe (2003) as V=15 Be star. Subsequently Cowley \& Schmidtke (2003)
analysed the long term light curve of this object obtained from the
MACHO data base. They showed that the optical object exhibited
outbursts every 394d which they proposed to be the binary period of
the system. Furthermore they also showed the presence of a
quasi-periodic modulation with a period $\sim$11d which they
associated with the rotation of the Be star's disk.

The MACHO red band data are reproduced in the top panel of 
Figure 1 for a period of
approximately 4 years. The outbursts reported by Cowley \& Schmidtke
(2003) are very clear at the interval of
394d. From the timing of the outbursts one can define a ephemeris of:

T$_{outburst}$ = JD2449830 + 394N\\
These epochs of visible outbursts are marked on Figure 1 with the ``V''
symbol.

The X-ray source has certainly been detected 5 times to date, 3 times by the
ASCA satellite observatory (Yokogawa et al, 2000) and twice by the
Rossi XTE spacecraft (Laycock et al, 2003). The two RXTE detections
are indicated on Figure 1 by an ``X'' symbol. The phase of these two
detections is exactly synchronised with the above ephemeris derived
from the optical outbursts. Therefore there can be no further doubt
that this period of 394d represents the binary period of the system,
with X-ray outbursts synchronised with the periastron passage of the
neutron star (NS). However, the ASCA X-ray outbursts are harder to
integrate into this model. They are indicated on Figure 1 by  ``A''
symbols. It is immediately clear that the ASCA X-ray outbursts have
occurred at different times to the MACHO/RXTE ephemeris and raises the
question as to whether it could be a different source entirely (see
Discussion).

\begin{figure*}
\psfig{file=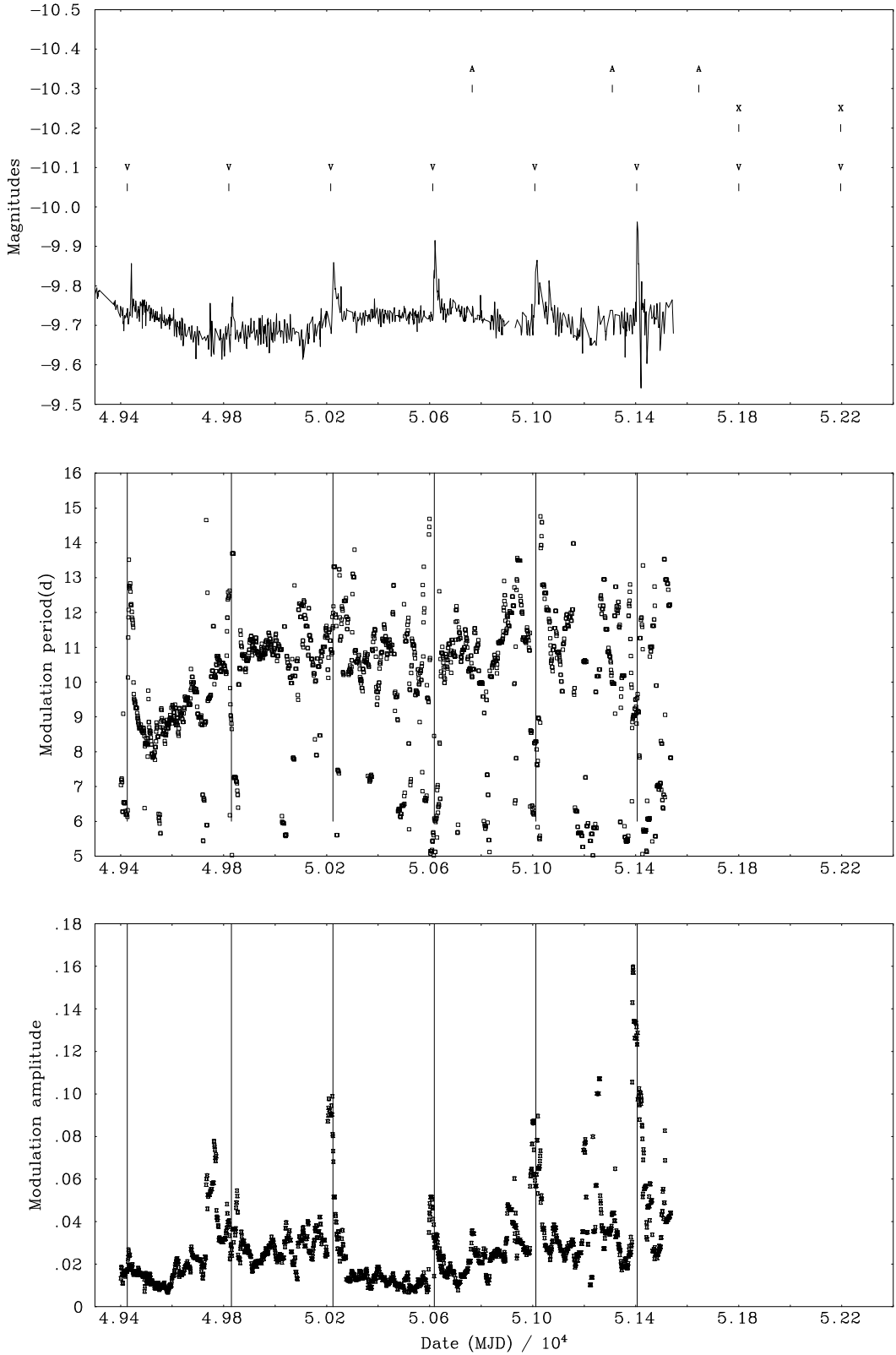,width=5.5in}
\caption{Top panel : the MACHO red lightcurve for the system AX J0049.4-7323 over
approximately 4 years. The ``V''
symbols indicate the phase of the optical outbursts, the ``X'' symbol the
epoch of the RXTE outbursts and the ``A'' sysmbols the epoch of X-ray
outbursts seen by ASCA (see text for references).
Centre panel : the measured period of the modulation determined by
studying a moving sample of length 80 days. Bottom panel : the
amplitude of the modulation using the same time bins as in
the central panel. The amplitude corresponds to the magnitude of
the most significant peak in the power spectrum.}\label{fig:lc}
\end{figure*}


\subsection{The $\sim$11d modulation}

The reported $\sim$11d modulation in the MACHO data for AX
J0049.4-7323 was investigated in detail here. Cowley \& Schmidtke
(2003) suggest that this arises from the interaction of the NS with
the circumstellar disk of the Be star. The MACHO data were analysed in
moving blocks of 80d to see how the period and amplitude varied over
the whole data set. Each block of 80d was subjected to a Lomb-Scargle
analysis and the position and amplitude of the highest peak in the
power spectrum
determined. The data sample was then moved on by 1d and the process
repeated. The results are presented in the lower two panels of Figure
1 (each 80d block result is displayed at the mid point of the time
interval being investigated). From this it can immediately be seen
that there is a strong increase in the pulse amplitude associated with
the 394d cycle peaking at optical outburst. Furthermore the modulation
period is also often substantially disturbed at the same epochs. The
occasional points at 6-7d represent times when the power spectrum
becomes more complex - see Figure 2 for two examples of the power
spectra from individual 80d samples.

\begin{figure}
\psfig{file=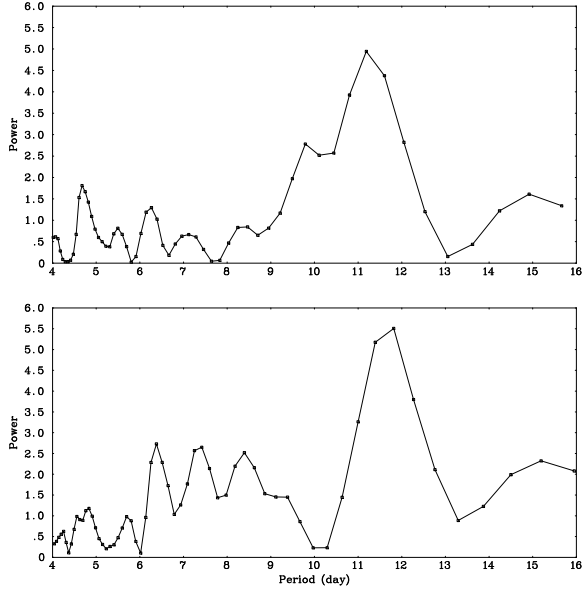,angle=0,width=3in}
\caption{Two examples of the Lomb-Scargle power spectrum obtained from
individual 80d samples. The upper example shows one of the periods in
which the $\sim$11d period is very strong. The lower curve shows one
of the less frequent intervals in which the power spectrum becomes
more complex and other frequencies share the power.
}\label{fig:ls}
\end{figure}

\section{Other similar systems}

The observed optical modulation in AX J0049.4-7323 is not unique in
the published literature. There are at least three further systems in
which a similar optical modulation is observed. These are listed in
Table 1.

In each case the MACHO lightcurve in the red band was extracted from
the archive and the data folded at the binary period. The resulting
folded lightcurves are presented in Figure 3. 

\begin{figure}
\psfig{file=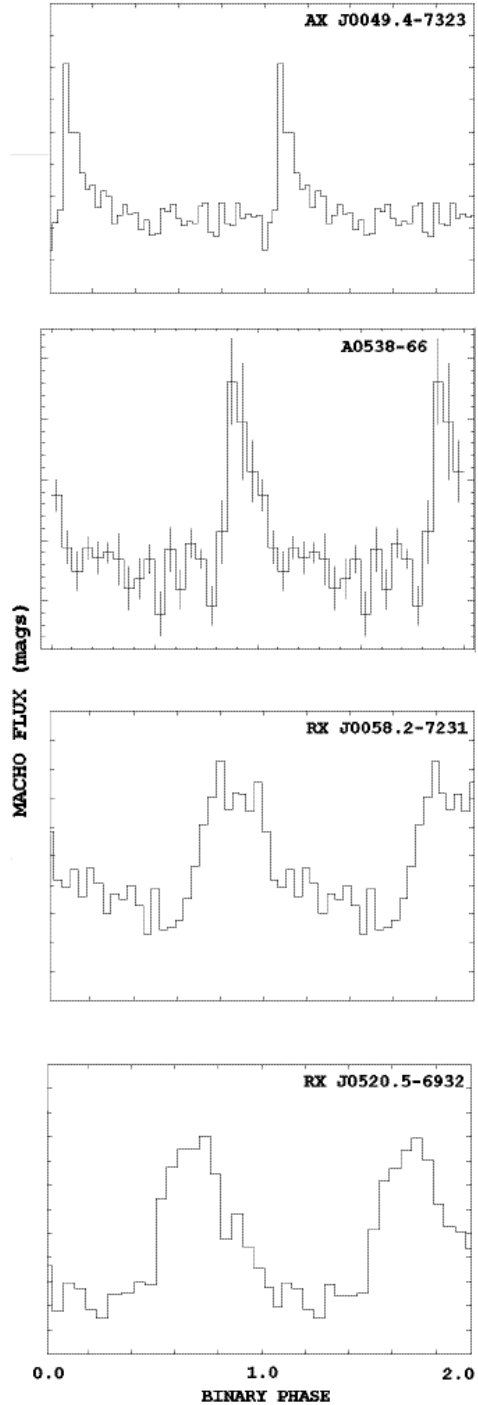,angle=0,width=2.5in}
\caption{The MACHO optical lightcurve folded modulo the binary period
of each system. Except for AX J0049.4-7323 the phase is arbitary. 
In each case the folded lightcurve is shown twice for
clarity. See Table 1 for details. The A0538-66 figure is taken from
Alcock et al, 2001.
}\label{fig:all4B}
\end{figure}

The lightcurves are all very assymetric, and show strong
evidence for a sharp rise followed by a slower decline. This effect is
clear in all the lightcurves shown, even though the binary periods
range from 394d to 16d (AX J0049.4-7323 to A0538-66).

\begin{table}
\centering
\begin{tabular}{|l|c|c|l|}
\hline
Object &Pulse &Orbital&References\\
Name&per(s)&per(d)&\\
\hline
AX J0049.4-7323&755&394&Cowley \& \\
&&&Schmidtke (2003)\\
A0538-66&0.069&16.6&Alcock et al., 2001\\
RX J0058.2-7231&?&59&Schmidtke, Cowley \\ 
&&&\& Levenson, 2003\\
RX J0520.5-6932&?&24.5&Coe et al, 2001\\
\hline
\end{tabular} 
\caption{List of periods observed in the systems.}
\end{table}

\section{Discussion}

\subsection{Does AX J0049.4-7323 = XTE J0049-723?}

The probable association of the ASCA X-ray source with the MACHO/RXTE
X-ray source seems quite high. The ASCA error circle (illustrated in
Figure 1 of Edge \& Coe, 2003) has a radius of $\sim$1 arcmin and
includes the MACHO object (albeit just inside the Eastern edge). In
addition, the reported ASCA pulse period of 755.5s is very close to
the RXTE period of 751s. 

The pulsar spin rate between the last ASCA detection and the first
RXTE detection some 5 months later can be used to estimate the average
X-ray luminosity. Using Equation 6.16 from Frank, King \& Raine (1992)
one can determine that such a change in spin rate over such a period
of time requires an average X-ray luminosity of $L_{x} = 3.10^{36}$
erg/s using a value for $\mu_{30}$ of 1. However, values of
$\mu_{30}$ in the range 0.1-10 produce results in the 
range $L_{x} = (0.1 - 4.0).10^{36}$ erg/s.
Laycock et al (2003) estimate
the detected X-ray luminosity in the outbursts to be $L_{x} =
2.10^{37}$ erg/s, whereas the ASCA detected luminosities are estimated
by Yokogawa et al (2000) to be only $L_{x} = 5.10^{35}$ erg/s (well
below the RXTE detection threshold). 
Therefore depending upon the correct value of $\mu_{30}$ it could well
be possible for the pulsar to have spun up over the $\sim$5 months.
Laycock et al (2003) carried out eight RXTE
observations of the SMC which included (but did not detect) this
source between the last ASCA detection and the first RXTE
detection. However, the source may well have been just below its
sensitivity limit.

The timing of the ASCA X-ray outbursts presents a puzzle. They
do not coincide with the phase of the optical and XTE outbursts. For
these kind of systems one normally sees two kinds of X-ray outbursts:\\ 

- Type I outbursts associated with the periastron passage of the NS
through the circumstellar disk\\

- Type II which occur randomly in the binary phase and are thought to
be associated with an unusually large mass ejection event from the
mass donor star. They often last in excess of a binary orbit implying
that the mass outflow is flooding the entire orbital space.

Not surprisingly, Type II outbursts are normally substantially more
X-ray luminous than Type I. The ASCA outbursts reported by Yokogawa et
al (2000) do not fit into either category - they are not at the right
phase for Type I, nor are they bright enough and long enough for Type
II. The intervals between the ASCA outbursts are 544d and
336d. Therefore the only other possibility is that the source spent an
extended period of time showing persistent, low-level X-ray emission
similar to X Persei and other supergiant systems. However, there are
no systems known to exhibit this mixture of persistent
and outburst characteristics, so this is hard to understand.

Could the ASCA and RXTE objects be different systems but with a
coincidentally similar pulse period? There is certainly the salutary
example of 4U 1145-619 and 1E 1145-619, pulse periods of 292s and 297s
respectively, and lying just 15 arcmin apart in the galaxy (Lamb et
al, 1980). These objects caused much confusion with early spacecraft
observations until it was realised they were two separate systems. In
addition, in the case here of AX J0049.4-7323, Laycock et al (2003)
point out that the reported pulse profiles between RXTE and ASCA are
very different from each other. Perhaps only better observations with
imaging X-ray telescopes will resolve this matter.

\subsection{The $\sim$11d modulation in AX J0049.4-7323}

\begin{figure}
\psfig{file=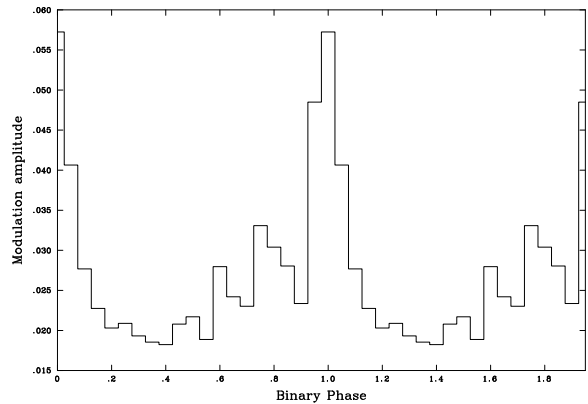,width=3in,angle=-90}
\caption{The amplitude of the $\sim$11d modulation folded modulo the
394d binary period. Phase zero is defined in Section 2.1 as the time
of peak optical outburst. The data are shown twice for clarity.}
\label{fig:fold}
\end{figure}

Both of the lower curves shown in Figure 1 strongly suggest the idea
that the $\sim$11d modulation is induced by the periastron passage of
the NS. If the modulation amplitude is folded modulo the amplitude and
phase of the binary modulation given in Section 2.1, then a strong
regular pattern is revealed - see Figure 4. Interestingly there is
little evidence to show the induced modulation systematically
decreasing in the intervals between NS passages.  On the contrary,
there is some evidence in this figure for the amplitude building up
over the $\sim$180d prior to the optical outburst. The smaller spikes
seen in the lower panel of Figure 1 at different phases than the
outburst phase can be seen contributing to the average curve shown in
Figure 3 around phases 0.6-0.8. All this activity well before phases
0.0 strongly suggests that the disk feels the effect of the approaching NS
long before the force becomes so strong that the disk can spill over
from its previously confined configuration (and hence enhance the
optical luminosity).

Similar psuedo periodic modulations have been reported by Schmidtke et
al (2004) from MACHO data of 2-3 other Be/X-ray systems in the
Magellanic Clouds. So this phenomenum may exist in many of these
systems and comparison of the differences and similarites in the
behaviour patterns may well prove extremely valuable in understanding
the neutron star/circumstellar disk interactions.

\subsection{The binary modulation lightcurve}

The first obvious feature to note from these light curves is that they
are not representative of the lightcurve one obtains from the extended
envelope of a star in a tight binary system. Such a situation will
produce a very sinusoidal modulation such as that seen in RX
J0050.7-7316 and modelled by Coe \& Orosz (2000).

Okazaki \& Negueruela (2001) suggest that in these systems the
circumstellar disk is truncated inside the periastron separation, and
hence the optical outburst would arise from the previously stable
configuration being disrupted at periastron. If this model is correct
then the different profiles may well represent differing inclination
angles of the NS to the circumstellar disk. Hence the relative rate of
passage past the disk would give rise to narrower or broader profiles
- an effect independent of the binary period.
Interestingly, the outburst profile is much sharper for the two
systems with known pulse periods (AX J0049.4-7323 and A 0538-66) than
the other two. This suggests a possible link between the system
orientation to the line of sight and our ability to see the poles of
the NS.

\section{Conclusions}

Detailed evidence on the system AX J0049.4-7323 has been presented
here to show how the passage of the neutron star disrupts the
circumstellar disk of the Be star. A similar effect is noted in three
other Be/X-ray binary systems. Together the observational data should
provide valuable tools for modelling these complex interactions.

\section*{Acknowledgments}

This paper utilizes public domain data obtained by the MACHO Project,
jointly funded by the US Department of Energy through the University
of California, Lawrence Livermore National Laboratory under contract
No. W-7405-Eng-48, by the National Science Foundation through the
Center for Particle Astrophysics of the University of California under
cooperative agreement AST-8809616, and by the Mount Stromlo and Siding
Spring Observatory, part of the Australian National University.

We are grateful to an anonymous referee for helpful comments that
significantly improved this paper.

\end{document}